\documentclass[%
superscriptaddress,
aps,
onecolumn
preprint,
]{revtex4-2}

\usepackage{graphicx}
\usepackage{dcolumn}
\usepackage{bm}
\usepackage{amsmath}
\usepackage{amssymb}

\begin{document}

\title{How turbulence increases the bubble-particle collision rate}

\author{Linfeng Jiang}
\affiliation{Physics of Fluids Group, Max Planck Center Twente for Complex Fluid Dynamics and Johannes Martinus Burgers Centre for Fluid Dynamics, University of Twente, 7500 AE Enschede}			

\author{Dominik Krug}
\email[]{d.j.krug@utwente.nl}
\affiliation{Physics of Fluids Group, Max Planck Center Twente for Complex Fluid Dynamics and Johannes Martinus Burgers Centre for Fluid Dynamics, University of Twente, 7500 AE Enschede}
		
\date{\today}

\begin{abstract}
We study the effect of turbulence on collisions between a finite-size bubble and small inertial particles based on interface-resolved simulations. Our results show that the interaction with the flow field around the bubble remains the dominant effect. Nonlinear dependencies in this process can enhance the turbulent collision rate by up to 100\% compared to quiescent flow. Fluctuations in the bubble slip velocity during the interaction with the particle additionally increase the collision rate. We present a frozen-turbulence model that captures the relevant effects providing a physically consistent framework to model collisions of small inertial particles with finite-sized objects in turbulence. 
\end{abstract}

\maketitle

\section{Introduction}
\label{sec:introduction}
Bubble-particle collisions are central to the flotation process, which is widely used e.g. in the mining industry to separate minerals through their attachment to rising bubbles. This intricate process involves a wide range of colloid science disciplines \citep{kostoglou_critical_2020,Nguyen2004,Derjaguin1993} with significant complexity stemming from the interplay of hydrodynamics and physicochemical interactions. The hydrodynamic bubble-particle interaction remains of paramount importance for optimization, as it is the rate-determining step that also sets the conditions for other interactions. 

Most of the work on bubble-particle collisions draws on the large body of work on particle-particle collisions \citep{pumir_collisional_2016}, in particular the seminal work of \citep{saffman_collision_1956} for the tracer limit and the model proposed by \citep{abrahamson_collision_1975} for the kinetic gas limit at large particle inertia. 
Such models predict the collision rate without considering the hydrodynamic interaction between the collision partners, but must take into account effects such as the segregation of bubbles and particles in a turbulent flow\citep{chan_bubbleparticle_2023}. 
However, the significant size difference between the larger bubble and the mineral particles in a typical flotation process, makes it important to account for the flow distortion around the bubble. 
This can cause an encountering particle to be deflected, reducing the actual number of collisions. 
For a bubble rising in quiescent flow, this process is deterministic and has been widely studied \citep{schulze_hydrodynamics_1989,Nguyen2004, sarrot_determination_2005, huang_effect_2012}. In this case, the effect of the hydrodynamic interaction can be captured in terms of collision efficiency, which relates the rate of actual collisions to the encounter rate. The lack of a suitable concept to account for turbulence in this context is widely acknowledged in the literature \citep{pyke_bubble_2003,nguyen_review_2016, Soldati2021}. Existing approaches are flawed because they combine turbulent encounter rates with collision efficiencies applicable to pure gravitational settling \citep{Bloom2002,Koh2007,Yoon2016, Liu2009}. Others use a Reynolds-type decomposition, which is problematic given the strongly nonlinear dependence of the problem on the flow velocity. Another conceptual inconsistency is that the encounter rate is based on the bubble-particle relative velocity, whereas the collision efficiency is determined by the bubble slip relative to the fluid. Some of these deficiencies are overcome \citep{kostoglou_generalized_2020}, which however remains limited to tracer particles.

The core problem is to determine the collision rate of small inertial particles with a finite-size  object in a turbulent flow. This is of general relevance to many other applications beyond flotation, such as the collision of cloud droplets \citep{falkovich_2002,Poydenot_Andreotti_2024}, the accretion of planetesimals by the collection of dust particles \citep{Guillot2014,Homann2016}, depth filtration \citep{cushing1998depth,May1967}, or bacterial degradation of marine snow \citep{ArguedasLeiva2022}. However, the scarcity of data due to the difficulties in performing the necessary experiments and interface-resolved simulations, has hindered the progress of theoretical research in this area. 
In this paper, we present a combined numerical and theoretical study of the effects of turbulence on bubble-particle collisions using direct numerical simulations.  We adopt homogeneous and isotropic turbulence (HIT), in which we measure how  turbulent fluctuation modifies the bubble-particle collision rate along the bubble-rising path. 

\section{Statistical Model}
\begin{figure}
	   \begin{center}
		\includegraphics[width=1\columnwidth]{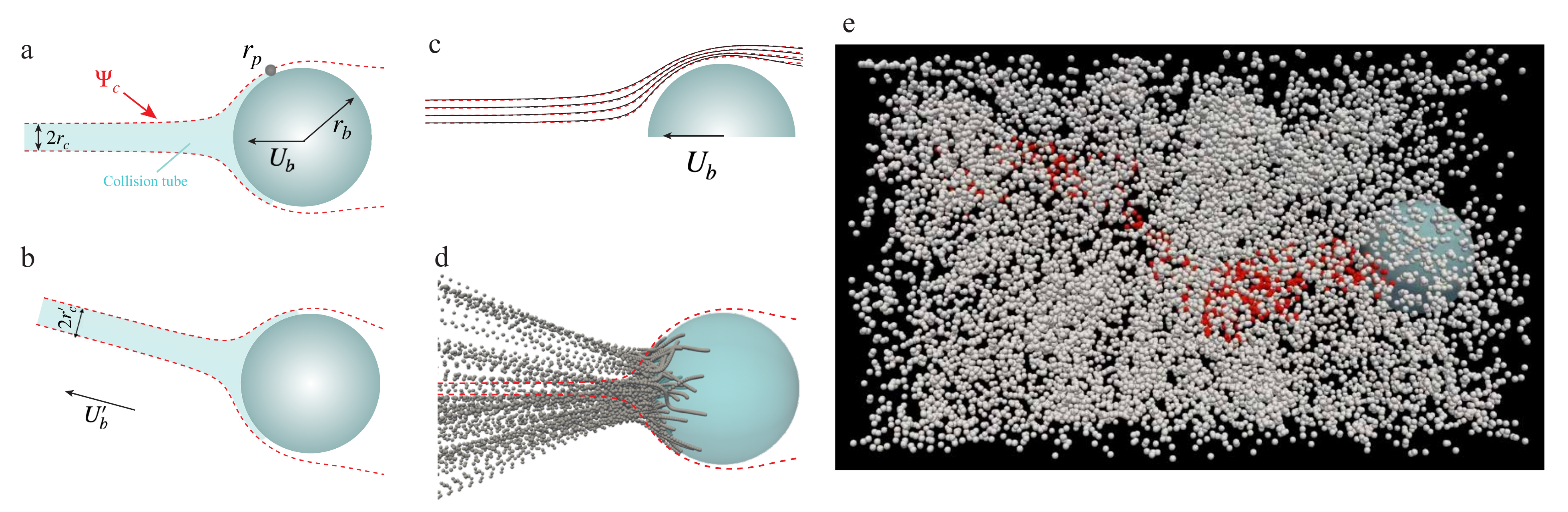}	
		\caption{(a) Sketch of the grazing trajectory (red dashed line) in quiescent flow. The shaded region indicates the collision tube where all particles collide on the bubble.
   (b) Sketch of the bubble-particle collision model under temporary bubble slip velocity $U_b'$.
   (c) Mean flow streamlines around the bubble for the case of imposed velocity bubble (solid) and quiescent flow (dashed lines) at $\overline{Re_b}=120$. (d) Trajectories of colliding particles ($r_p/r_b=0.05,~St_p=0.04$) for the case of imposed velocity bubble compared to the corresponding grazing trajectories (red lines) in quiescent flow at the same $Re_b =120$. (e) Snapshot of the bubble-particle collision process in turbulence for the imposed-velocity bubble with flow from left to right in the bubble frame of reference. Incoming particles that end up colliding with the bubble are marked in red.
		}
		\label{fig:fig_visualization}			
    \end{center}
\end{figure}

The collision frequency $K$ of particles with a bubble is determined by $K = \Gamma n_p$, where $n_p$ is the particle number density, and the proportionality coefficient $\Gamma$ ($m^3s^{-1}$) is commonly referred to as the collision kernel and is determined by the flow \citep{pumir_collisional_2016}.
In quiescent flow, the collisions between a spherical bubble of radius $r_b$ rising at constant velocity $U_b$ with small inertial solid particles are deterministic. The particle grazing trajectory $\Psi_c$ determines a 'collision tube', encompassing all the particles contained inside collide with the bubble, as illustrated in Fig.~\ref{fig:fig_visualization}(a). In this case, the collision frequency is determined by the collision number flux: $K=Q_cn_p$. The collision number flux $Q_c$ can be determined by the cross-section area of radius $r_c$ limited by the grazing trajectories upstream far from the bubble: $Q_c=\pi r_c^2U_b$.  
Therefore, the collision kernel in quiescent flow can be expressed as $\Gamma_q=Q_c=\pi r_b^2U_bE_c$, where the collision efficiency $E_c=r_c^2/r_b^2$ measures the ratio of collided particle number to the total encountering particle number. 
Notice here $E_c$ depends on the bubble Reynolds number $Re_b=2r_bU_b/\nu$, the ratio of particle radius $r_p$ and $r_b$, and the Stokes number $St_p = \tau_p/\tau_f$, with $\tau_p$ the particle response time and $\tau_f = 2r_b/U_b$ the timescale of the bubble-particle interaction \citep{dai_review_2000, sarrot_determination_2005,huang_effect_2012}.

To provide an understanding of the relevant physical mechanisms of bubble-particle collision in turbulence, we propose a statistical model. 
We start from the assumption that the flow field in the vicinity of the bubble is approximately stationary and uniform during the bubble-particle interaction. Conceptually, this implies that the temporal scale $\tau_f$ is (much) shorter than the correlation timescale of flow fluctuations ($O(\tau_\eta)$), but it remains to be tested from the data under what conditions exactly this assumption is valid in practice.
Additionally, we consider the flow correlation length scale to be comparable to or larger than the bubble size. Within this `frozen turbulence' assumption, the instantaneous bubble-particle collision process in turbulence can be related back to that observed in quiescent flow. Therefore, the equivalent steady flow problem is characterized by the magnitude of the instantaneous bubble slip velocity $U_b'$ with corresponding values of $Re_b' = 2r_bU_b'/\nu$ and $St_p'=\tau_p/(2r_b/U_b')$, as shown in Fig.~\ref{fig:fig_visualization}(b).
The entire bubble-particle collision process in a turbulent flow can then be viewed as a superposition of collision events in quiescent flow with varying parameters $Re_b'$ and $St_p'$. In the framework, the collision number ${N_c|_{\Delta\tau}}$ during a short period $\Delta \tau$ ($O(\tau_\eta)$) in turbulent flow, with mean particle number density $n_p$, is given by $N_c|_{\Delta\tau}=\pi r_b^2E_c(Re_b',St_p')U_b'n'_p\Delta\tau$,
where $n_p'$ is the instantaneous encountering particle number density, which depend on $Re_b'$ and $St_p'$. Importantly, $E_c(Re_b',~St_p')$ denotes the collision efficiency in quiescent flow for varying flow parameters, which is deterministic and can be parameterized. Then, the total expected collision number $N_c$ during a sufficient long time period $T=\sum \Delta \tau$ is given by 
\begin{equation}
	N_c = \pi r_b^2T\int E_c(Re_b',~St_p')U_b'f(U_b')n_p'\rm d \it U_b', \label{eq:N_c'}
\end{equation}
Based on this, the collision kernel can be derived as follows:
\begin{equation}
	\Gamma = \frac{K}{n_p}={\frac{\pi r_b^2}{n_p}\int E_c(Re_b',~St_p')U_b'f(U_b')n_p'{\rm d} U_b'}. \label{eq:E_c model}
\end{equation}
Note that collision efficiency has been extensively studied, and the general dependence of $E_c$ on the bubble Reynolds number and particle Stokes number has been established in the literature \citep{schulze_hydrodynamics_1989,saffman_collision_1956, huang_effect_2012}.
The collision kernel is thus determined by the probability density function (PDF) $f(U_b')$ of the bubble slip velocity $U_b'$, which remains an open question. 
Furthermore, we note that the collision kernel will also be affected by the preferential sampling for particles with large $St_p$ as this this alters the incoming particle number density.

\section{Numerical methods and simulations}
We perform interface-resolved direct numerical simulations to test this model. The turbulent flow is governed by the incompressible Navier-Stokes equations (NSE), which read:
\begin{eqnarray}
		\frac{\partial\textbf{u}}{\partial t}  + \textbf{u}\cdot {\nabla}   \textbf{u} &=& -   \frac{1}{\rho_f}{\nabla} p + \nu\ \nabla^2   \textbf{u} + \textbf{f} + \textbf{f}_b, \label{eq:N-S}\\
		{\nabla}  \cdot  \textbf{u} &=& 0, \label{eq:div}
\end{eqnarray}
where $\mathbf{u}$ is the fluid velocity, $p$ denotes the pressure and the parameters are the kinematic viscosity $\nu$ and the reference liquid density $\rho_f$. The vector $\mathbf{f}$ denotes an external random large-scale volume force, which is statistically homogeneous and isotropic, with constant-in-time global energy input \citep{Perlekar2012}. This force is used to generate and maintain the turbulent flow. The bubble-free turbulent flow intensity is characterized by the Reynolds number based on the Taylor microscale, $Re_\lambda=\sqrt{15u'/(\nu\epsilon)}$, where $u'$ denotes the root-mean-square velocity of the turbulence and $\epsilon$ is the mean energy dissipation rate. The vector $\mathbf{f}_b$ accounts for the bubble-fluid two-way coupling.

In the practice of flotation, commonly a large amount of surfactants is present in the liquid. Therefore, it is reasonable to assume that the bubbles with a typical Reynolds number below 200 are fully contaminated and approximately spherical, leading to a nearly no-slip boundary condition \citep{Nguyen2004, huang_effect_2012}. The Weber number $We=2r_b\rho_f U_b^2/\chi $ based on the surface tension $\chi$ of water and the bubble rise velocity is O(0.1) for our simulations and even lower based on turbulent fluctuations. Therefore bubble deformations remain negligible even if surface tension is lowered by the presence of surfactants. Under these conditions, the bubble behaves similarly to buoyant spheres. 
In this case, the translation and rotation of the bubble are governed by the Newton-Euler equations (NEE)
\begin{eqnarray}
		m_b \frac{\rm{d} \mathbf{U}_b}{{\rm d}t} &=& \oint_{S_b} \bm{\sigma} \cdot \mathbf{n}\ {\rm d}S + V_b(\rho_b-\rho_f)\mathbf{g}, \label{eq:Newton-Euler1}\\
		\frac{{\rm d}  \bm{\mathbf{I}} \mathbf{\Omega}}{{\rm d}t} &=& \oint_{S_b} (\mathbf{x}-\mathbf{x}_b) \times  (\bm{\sigma} \cdot \textbf{n})\ {\rm d}S, \label{eq:Newton-Euler2}
\end{eqnarray}
where $\mathbf{U}_b(t)={\rm d}\mathbf{x}_b/{\rm d}t$ and $\mathbf{\Omega}(t)$ respectively are the bubble velocity and angular velocity vectors at position $\mathbf{x}_b(t)$. The bubble mass is given by $m_b=\rho_b V_b$ ($\rho_b$ the bubble density and $V_b$ the volume) and $\bm{\mathbf{I}}$ is the moment of inertia tensor.
$\bm{\sigma} = -p \bm{I}+\rho \nu (\nabla\textbf{u} +  \nabla\textbf{u}^T)$ is the fluid stress tensor, $\textbf{x}-\textbf{x}_b$ the position vector relative to the bubble center and $\textbf{n}$ the outward-pointing normal to the bubble surface $S_b$.
This term $\bm{\sigma}$ is solved by the immersed boundary method (IBM) to account for the two-way coupling between the flow and the bubble. In the last decades, IBM has been widely used for studying the multi-phase flow \citep{Peskin2002, Mittal2005, Uhlmann2005}. In IBM, the Euler grids of the fluid are fixed and the surface of the bubble is represented by Lagrangian nodes which move with the bubble motion. 
To avoid the formation of a mean flow due to the buoyancy driving exerted by the bubble, 

we compensate the average force applied from the bubble to the liquid to attain a statistically stationary state \citep{Hoefler2000, Chouippe2015}. To achieve this, the spatial average of the IBM volume force term needs to be subtracted. More explicitly, we compute the average at each simulation time step:
\begin{eqnarray}
	\langle \mathbf{f}^{ibm}\rangle_\Omega(t) = \frac{1}{||\Omega||}\int_\Omega \mathbf{f}^{ibm}(\mathbf{x},~t) \rm{d}\mathbf{x}, \label{eq:mean_ibm_force}
\end{eqnarray}
where $\Omega$ denotes the entire computational domain, $\langle \cdots\rangle_\Omega$ indicates the spatial average over the entire $\Omega$ region and $||\Omega||$ is the volume of $\Omega$ region. Then, the bubble-related contribution to the volume force, $\mathbf{f}_b$, is obtained by 
\begin{eqnarray}
	\mathbf{f}_b(\mathbf{x},~t)=\mathbf{f}^{ibm}(\mathbf{x}~,t)-\langle \mathbf{f}^{ibm}\rangle_\Omega(t), \label{eq:compens_ibm}
\end{eqnarray}

We adopt an implicit method \citep{Tschisgale2017} to solve the bubble dynamics as the conventional explicit method is numerically unstable when the bubble is light \citep{Uhlmann2005}.
To avoid the built-up of momentum in the triply periodic domain over time, the force applied by the bubble on the liquid is compensated to attain a statistically stationary state \citep{Chouippe2015}.

\begin{table}
	\begin{center}
			\begin{tabular}{p{3.4cm}p{1.1cm}p{1.1cm}p{1.1cm}p{1.1cm}p{1.1cm}p{1.1cm}p{1.1cm}p{1.1cm}p{1.1cm}}
			Simulation cases	& $Re_\lambda$ & $\eta/\Delta x$ & $\tau_\eta/\Delta t$ & $L/\eta$ & $T_L/\tau_\eta$ & $\lambda/\eta$   & $\overline{Re_b}$  & $T_i$\\
			 imposed-velocity bubble &	32 &  8 & 780 & 27.3 & 9.1 & 11.6 & 120 & 0.25 \\
             freely rising bubble    &  32 &  8 & 780 & 27.3 & 9.1 & 11.6 & 110 & 0.27 \\
             freely rising bubble    &  64 &  4 & 780 & 67.9 & 16.7 & 16.2 & 150 & 0.53 \\
			\end{tabular}
			\caption{Parameter of the numerical simulations and relevant turbulence scales. $Re_\lambda$: Taylor-Reynolds number, $\eta=(\nu^3/\epsilon)^{1/4}$: Kolmogorov dissipation length scale in grid space units $\Delta x$, $\tau_\eta$: Kolmogorov time scale in time-step units $\Delta t$, $L=u'^3/\epsilon$: integral scale, $T_L=L/u'$: large-eddy turnover time, $\lambda = u'\sqrt{15 \nu / \epsilon }$: Taylor micro-scale, $Re_\lambda$: Taylor-Reynolds number, $\overline{Re_b}=2r_b\overline{U_b}/\nu$: bubble Reynolds number based on the mean rising velocity $\overline{U_b}$, $T_i=u'/\overline{U_b}$: turbulent intensity}
			\label{table:parameters}
	\end{center}
\end{table}

We use a code based on the lattice Boltzmann method (LBM) to solve the Navier-Stokes equations \citep{Calzavarini2019, Jiang2022}. 
Two sets of simulations are carried out: 1) a simplified ideal configuration where a constant bubble velocity is imposed, and 2) simulations with a freely rising bubble, where the bubble-fluid density ratio is $\rho_b/\rho_f=10^{-3}$. Matching practically relevant conditions \citep{wang_effect_2022}, we consider a moderately turbulent flow ($Re_\lambda=32$ and 64) and a bubble Reynolds number $Re_b\sim O(100)$. Detailed parameters of the simulations are summarized in table \ref{table:parameters}. 
There are two computational conditions that should be satisfied in the simulations. Firstly, the wake of the fully contaminated bubble presents a steady axis-symmetric vortex at such a high $Re_b$ in a quiescent flow \citep{sarrot_determination_2005,johnson_flow_1999}. As the flow domain is periodic, the incoming flow ahead the bubble might be disturbed by the remnants of this bubble wake. To avoid this issue, the flow domain in the bubble rising direction should be sufficiently large. The whole flow domain adopted in this work is rectangular with uniform grid sizes, $N_x \times N_y\times N_z=3072\times 512\times 512$, where the bubble rises in the $x$-direction. We found that using this domain length was sufficient to avoid wake effect. 
Secondly, the number of lattices within the boundary layer of the bubble should be sufficient. The approximate boundary layer thickness is estimated as $\delta=1.13/Re_b^{1/2}$ \citep{johnson_flow_1999}, where $\delta$ denotes the boundary layer thickness normalized by $2r_b$. Additionally, the minimum particle size ($r_p/r_b=0.025$) should be larger than the lattice unit to ensure the accuracy of interpolation when the particle is close to the bubble.
To this end, the bubble diameter is resolved by 80 grids to make sure the boundary layer is well resolved, as well as the minimum distance of the point-like particle to the bubble is larger than one grid unit. 
The ratio of turbulent dissipation time scale $\tau_\eta$ to $\tau_f$ in our simulations is close to 1.2.
We note that the typical auto-correlation time scale of the turbulent flow is longer than $\tau_\eta$, which implies the validity of the assumption of frozen turbulence in our model.
In the flotation process, the mineral particles are significantly smaller than the turbulent dissipation scale $\eta$. 
We represent these as one-way coupled point particles, considering the non-Stokesian drag force and the added mass force. 
We do not account for potential alterations to the drag force when a particle is close to the bubble surface.
The particle response time $\tau_p=r_p^2/(3\beta\nu)$ includes the density coefficient $\beta=3\rho_f/(\rho_f+2\rho_p)$, where $\rho_p$ denote the density of the particle. 
The particle dynamics is driven by the instantaneous turbulent flow. 
The collision frequency, $K=N_c/T$ and thus the collision kernel $\Gamma = K/n_p$, is measured by counting the number of particles ($N_c$) that collide with the bubble over a long time period $T$. 
A collision is detected when the distance between the particle and bubble is smaller than $r_p+r_b$.
Additionally, we conduct simulations of bubble-particle collisions in a quiescent flow at $Re_b$ ranging from 80 to 210 to obtain the dependence of $E_c$ on $Re_b$ and $St_p$. 
In these simulations, a constant-velocity inflow is applied at the inlet, while a homogeneous Neumann boundary condition is imposed at the outlet. The bubble is fixed at the center of the domain and the spatial resolution is kept the same as for the turbulent flow case and the turbulence forcing term $\mathbf{f}$ is switched off.

\section{Results}
We start from the discussion of the bubble with an imposed velocity. 
In Fig.~\ref{fig:fig_visualization}(c) we demonstrate that the impact of turbulent fluctuations on the mean streamlines is insignificant for the present parameters.
In particular in the incoming flow, which determines the collision rate, the differences are very small, lending support to our modelling approach. More noticeable differences in the wake region are a consequence of turbulence disrupting the symmetric recirculation pattern behind the bubble.
However, the influence of turbulence on the collision process is distinct, as becomes evident from the supplementary movie and Fig.~\ref{fig:fig_visualization}(d), where trajectories of colliding particles are shown.  These trajectories originate from a cone-shaped region that is notably larger than the corresponding collision radius $r_c$, based on the grazing trajectory in quiescent flow. 
Correspondingly, also the collision angle relative to the direction of the imposed bubble velocity is found to be wider in the turbulent flow (see Appendix \ref{appA}).
Nevertheless, the collision trajectories appear to follow straight paths towards the bubble, and the colliding particle stream exhibits a band-like pattern, as shown in Fig. \ref{fig:fig_visualization}(e), both of which are consistent with our modeling assumptions. 

\begin{figure}
	\begin{center}
		\includegraphics[width=0.8\columnwidth]{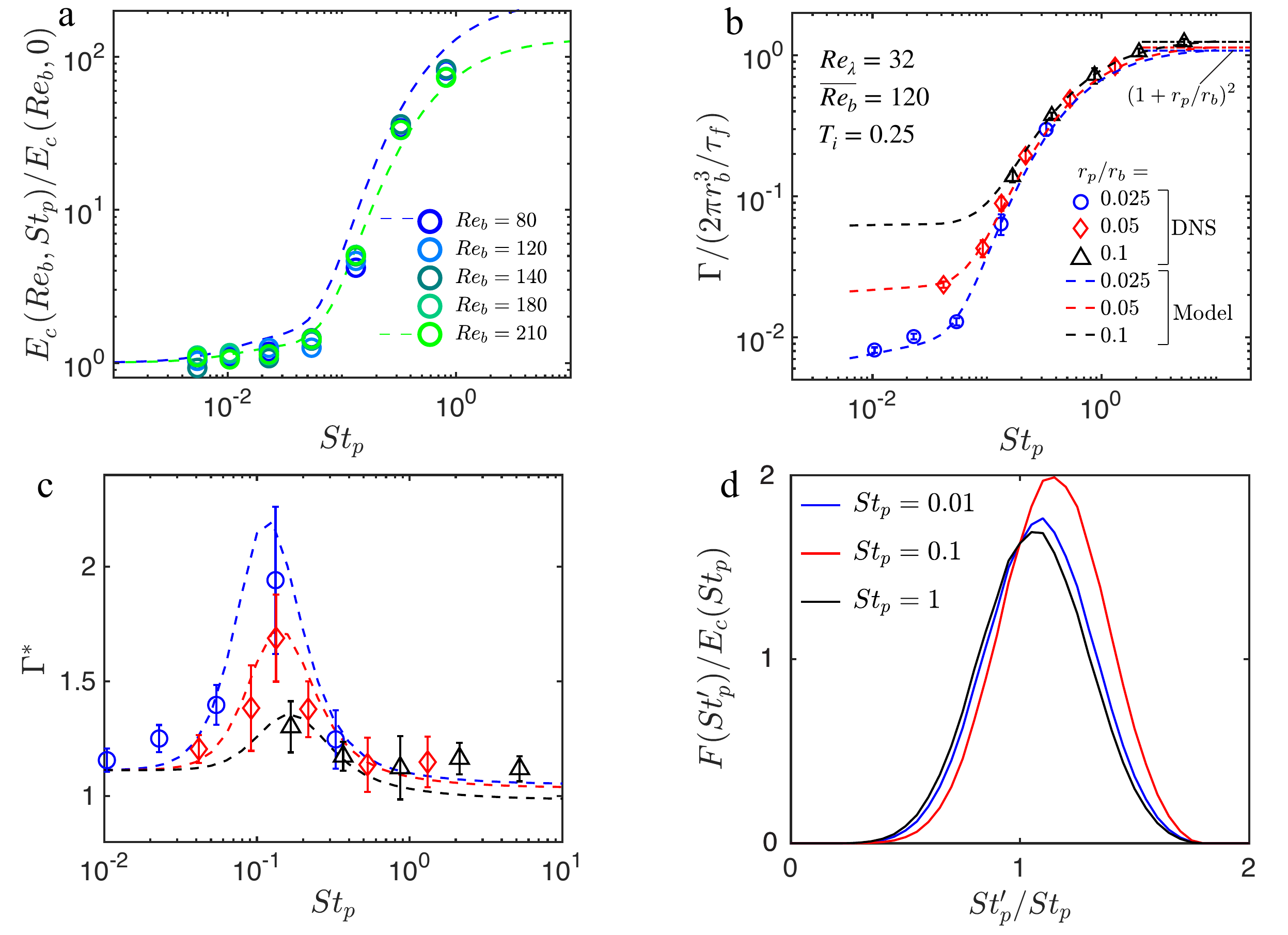}	
		\caption{(a) Normalised $E_c$ as a function of $St_p$ in quiescent flow for various $Re_b$ compared to the fits according to Eq.~(\ref{eq:Ec_Stp_schulze}) (dashed lines). (b) Dimensionless collision kernel vs. $St_p$ at $\overline{Re_b}=120$ for bubble with imposed velocity in HIT (symbols) and model (dashed lines). (c) Turbulent collision kernel relative to that in quiescent flow for the imposed velocity bubble. Error bars represent fluctuations between subsets of the data. (d) Scaled PDF of $St_p'$ as a function of $St_p'$ for different $St_p$.
		}
		\label{fig:fig_Ec_Stp}			
    \end{center}
\end{figure}

For quiescent flow, the dependence of $E_c$ normalised by the result for tracer particles $E_c(Re_b,0)$, on $Re_b$ and $St_p$ is presented in Fig.~\ref{fig:fig_Ec_Stp}(a).
For very small $St_p$, the inertial effect is negligible and the particles follow the flow streamlines, so that interception is the dominant factor determining the number of  collisions in this range \citep{dai_review_2000}. Analytical predictions based on flow streamlines in quiescent flow indicate that $\Gamma$ scales with $(r_p/r_b)^2$ \citep{weber_interceptional_1983} in this case.
However, as particle inertia becomes more pronounced, particles deviate from the flow streamlines and collide on the bubble even if their initial position is outside the grazing trajectory of the inertialess particle. This leads to a higher collision rate and explains the rapid increase in $E_c$ as $St_p$ approaches 0.1, beyond which the inertial effect dominates. At very large $St_p$, the particles are barely influenced by the flow such that $E_c$ approaches $(1+r_p/r_b)^2$. To be able to evaluate $E_c(Re_b, St_p)$ analytically, we employ an empirical expression, which is the sum of its two contributions: interceptional collision $E_i$ and the collision associated with particle inertia\citep{schulze_hydrodynamics_1989}:
\begin{equation}
    E_c = E_i+(1+\frac{r_p}{r_b})^2(\frac{St_p}{St_p+a})^b(\chi-\frac{E_i}{(1+r_p/r_b)^2}) \label{eq:Ec_Stp_schulze}
\end{equation}
Here, we adopt $E_i=\frac{3}{2}(r_p/r_b)^2(1+Re_b^{2/3}/5)$ \citep{sarrot_determination_2005} for the collision efficiency in the tracer limit and the fitting parameters are set to $a=0.2$ and $b=2$.  $\chi=1-0.9\cdot10^{-(\frac{log(St)+1.3}{1.6})^2}$ is a fitting correction term  to better capture the transition around $St_p\approx 0.1$. The resulting fit is in good agreement with the data for the evaluated parameter range as shown in Fig. \ref{fig:fig_Ec_Stp}(a).

As shown for the case with constant bubble velocity in Fig.~\ref{fig:fig_Ec_Stp}(b), the general trends of $\Gamma$ in turbulence, in particular the increase with increasing $St_p$ for all particle sizes are consistent with those observed in quiescent flow. 
In this simplified configuration, the PDF of the  $U_b'$ required to evaluate our model (Eq. \ref{eq:E_c model}), can be obtained by combining the constant bubble rise velocity with the Gaussian distribution of turbulent velocity fluctuations.
The collision kernel predicted in this way is in excellent  agreement with the simulations. This is further confirmed in Fig.~\ref{fig:fig_Ec_Stp}(c), where we scrutinize the result by plotting it relative to the collision kernel at the same bubble velocity in quiescent flow as $\Gamma^*=\Gamma/\Gamma^q$. In this way, it also becomes clear that turbulent flow significantly enhances the collision kernel. 
Here, two interesting aspects should be underscored:
Firstly, $\Gamma^*$ surpasses 1 when inertia is negligible ($St_p \ll 1$), indicating that turbulent fluctuations enhance interceptional bubble-particle collision.
Secondly, the collision enhancement is not uniform across the considered range of $St_p$, suggesting that the combined influence of turbulence and particle inertia leads to further amplification of the collision rate. The collision enhancement can reach approximately 100\% for particles with a size ratio of $r_p/r_b=0.025$. For larger particles, the maximum collision enhancement is lower, though the peak still occurs at a similar value of $St_p\approx 0.1$.

The good agreement with the simulations indicates that the model adequately captures the relevant turbulence effects on the collisions, enabling us to explore their origin. 
We notice that due to the increase of $E_c$ with increasing $Re_b'$, the integrand in Eq. (\ref{eq:E_c model}) nonlinearly depends on $U_b'$. This results in an increase in the predicted collision rate even if $f(U_b')$  is symmetric around $\overline{U_b}$. In the present case, this effect amounts to close to 15\% increase in $\Gamma$, consistent with what is observed at low $St_p$. 

In addition, there is an inertial effect as a change in $U_b'$ also changes $St_p'$. The strongly nonlinear dependence of $E_c$ on $St_p'$, especially in the intermediate range $St_p \approx 0.1$, leads to an asymmetric response to positive and negative velocity fluctuations.
This is demonstrated by the scaled PDF of $St_p'$, $F(St_p')=E_c(St_p')PDF(St_p')$, shown in Figure \ref{fig:fig_Ec_Stp}(d). For low and high $St_p$, the dependence of $F(St_p')$ on $St_p'$ is almost symmetric. However, for an intermediate value of $St_p\simeq0.1$,  the contribution to collisions from fluctuating $St_p'$ exhibits a positive bias, which explains the strongly enhanced turbulent collision rate in this range and the non-monotonic dependence on $St_p$. Due to the higher interceptional collision efficiency, the increase in the inertial range and hence the turbulent enhancement is less pronounced for larger size ratios $r_p/r_b$. 

\begin{figure}
	\begin{center}
		\includegraphics[width=1\columnwidth]{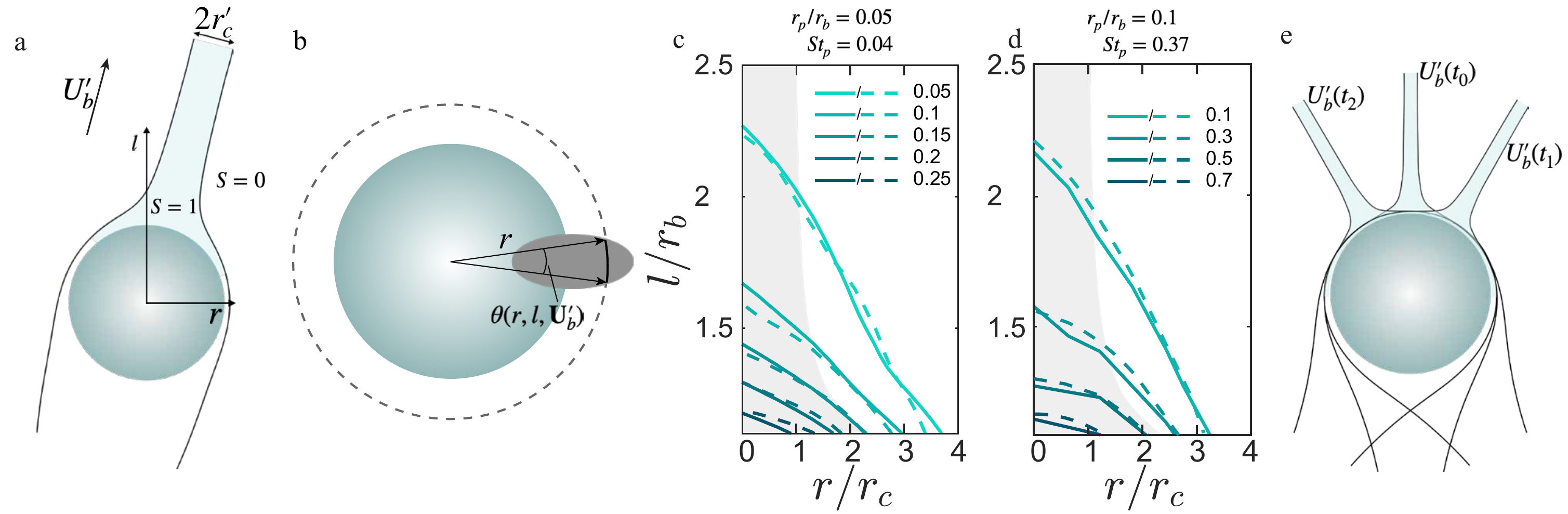}	
		\caption{(a) Sketch of the bubble-particle collision  with temporary slip velocity $\mathbf{U}_b'$. The blue shaded region indicates the binary function $S(r,l,\mathbf{U}_b')$, which is the projection of the collision tube on the $r-l$ plane. 
        (b) Sketch of the cross section (grey region) between the collision tube and the plane $l$ in the view along $l$ axis. $\theta(r,l,\mathbf{U}_b')$ indicates the radian of the arc that occupied by the grey region in the circle of radius $r$, which is used measure the term $G(r,l,\mathbf{U}_b')$.
        (c) and (d): Contour lines of $P(r, l)$ from simulations (solid) and model (dashed lines) for $(r_p/r_b=0.05, St_p=0.04)$ and $(r_p/r_b=0.1, St_p=0.37)$, respectively. 
        (d) Sketch of the collision probability under different bubble slip velocities. The region with more overlaps corresponds to the one with higher collision probability.
		}
		\label{fig:collision_probability}			
    \end{center}
\end{figure}

Another way to validate the model is to consider the spatial distribution of the collision probability $P(r,~l)$, where $r$ and $l$ are distances perpendicular and along the bubble velocity direction, respectively, (see Fig.~\ref{fig:collision_probability}(a)). 
In quiescent flow, the collision process is deterministic and the collisions probability is a binary function, which has a tube shape based on the grazing trajectory (see the shaded region in Fig.~\ref{fig:collision_probability}(a)). We represent this using the binary function $S(r,l,;\mathbf{U}_b)$, which is equal to 1 inside the collision tube and zero otherwise, as shown in Figure \ref{fig:collision_probability}(a). Practically, $S$ can be determined from the grazing trajectory.
Based on the `frozen turbulence' assumption, the collision probability in turbulence can be predicted as the superposition of the binary probability distributions corresponding to the instantaneous slip velocity $\mathbf{U}_b'$. This leads to
\begin{equation}
   P(r,l) = \int f(\mathbf{U}_b')S(r,l;\mathbf{U}_b')G(r,l,\mathbf{U}_b'){\rm d}\mathbf{U}_b'.
    \label{eq:Pmodel}
\end{equation}
Note that the additional factor $G(r,l,\mathbf{U}_b')=\theta(r,l,\mathbf{U}_b')/2\pi$ in Eq.~(\ref{eq:Pmodel}) is a geometrical coefficient related to the azimuthal integration required for the projection onto the two-dimensional $(r,l)$ space. It measures the fraction of the circle with radius $r$ that falls inside the collision tube at distance $l$ as illustrated in figure \ref{fig:collision_probability}(b).
The result of model prediction according to Eq.~(\ref{eq:Pmodel}) is again in excellent agreement with the data as shown in Fig.~\ref{fig:collision_probability}(c) and (d), confirming that the basic physical processes are well represented in the model. 
Consistent with Fig.~\ref{fig:fig_visualization}(d), the impact of turbulence is clearly evident leading to a much wider distribution of $P(r,~l)$ compared to the binary distribution in quiescent flow. 
This effect is especially pronounced if $St_p$ is low as is the case for Fig.~\ref{fig:collision_probability} (c). The collision efficiency is low and the associated collision stream tube slender for this case, such that fluctuations in the instantaneous bubble slip direction lead to low values of $P(r,~l)$ even close to the bubble surface, as is illustrated in Fig.~\ref{fig:collision_probability} (e). 
Since the collision efficiency increases for larger $St_p$ and the collision tube widens, this effect becomes less strong and the values of the collision probability close to the bubble are much higher in the plot for $St_p = 0.37$ in Fig.~\ref{fig:collision_probability} (d).

\begin{figure}
	\begin{center}
		\includegraphics[width=1\columnwidth]{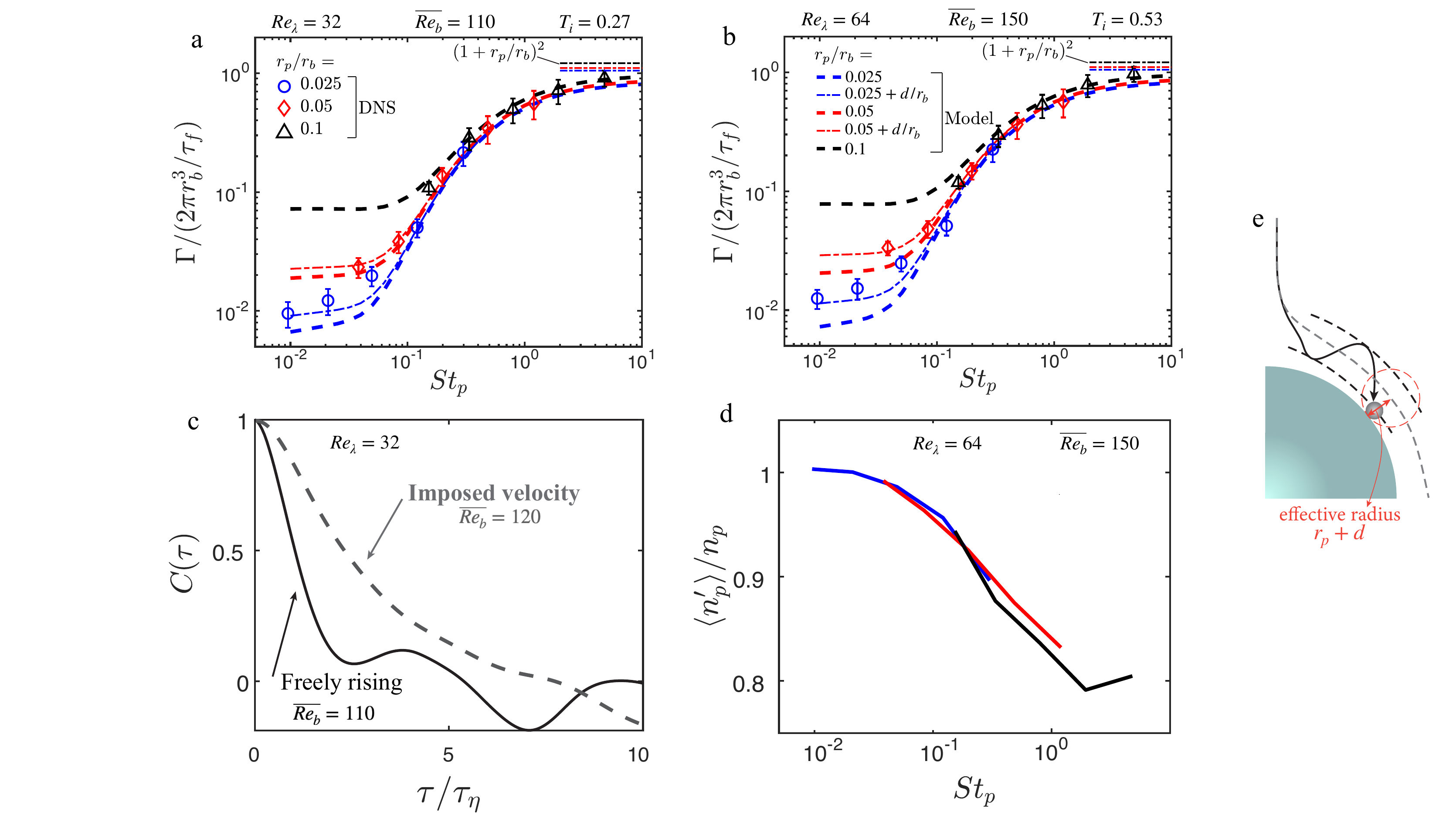}	
		\caption{
  Normalized collision kernel for the freely rising bubble at (a) $Re_\lambda=32$, $\overline{Re_b}=110$ and (b) $Re_\lambda=64$, $\overline{Re_b}=150$. 
  (c) Comparison of auto-correlation function of $U_b'$ for the imposed velocity ($Re_\lambda=32,~\overline{Re_b}=120$) and freely rising bubbles ($Re_\lambda=32,~\overline{Re_b}=110$), respectively. (d)) shows the normalized mean incoming particle number density as a function of $St_p$ at $Re_\lambda=64$. (e) Sketch in the bubble reference frame, showing how fluctuations $U_b'$ during the interaction effectively enlarge the particle collision radius. 
		}
		\label{fig:figure3}			
    \end{center}
\end{figure}

Having established the general suitability of the model to capture the relevant turbulence effects on the collision rate, we now turn to the more realistic case of a freely rising bubble. The corresponding results in terms of $\Gamma$ are shown in Fig.~\ref{fig:figure3}(a,b) for different $Re_\lambda$. In these cases, the bubble-slip velocity pdf $f(U_b')$ required as input for the model is measured by averaging the fluid velocity located on a spherical surface of radius $3r_b$ that is centred at the bubble's center position \citep{kidanemariam2013direct}. The size of the spherical surface is chosen in a way that the fluid velocity is not significantly influenced by the presence of the bubble boundary. The radius $3r_b$ of the spherical surface is tested in the case of uniform flow past a fixed sphere \citep{kidanemariam2013direct}, which results in a measured fluid velocity approximately $90\%$ of the incoming flow velocity.

The general agreement between $\Gamma$ predicted by the model (dashed lines in Fig.~\ref{fig:figure3}(a, b) and the data remains good also for the free rising cases. Difference arises at low $St_p$, where the model is found to underpredict the simulation result. This can be explained by the correlation time of $U_b'$, which is shorter for free rising bubbles compared to the imposed velocity case (see Fig.~\ref{fig:figure3}(c)). The resulting changes in $U_b'$ during the bubble-particle interaction cause the particle trajectory to fluctuate in the bubble frame of reference (see the sketch in Fig.~\ref{fig:figure3}(e)). This increases the effective collision radius of the particle to $r_p+d$, where $d$ is a measure of the drift from the original particle trajectory. The relevant time and velocity scales for this drift are $\tau_f$ and $u_\eta$, respectively, such that $d \sim \tau_f u_\eta$ in analogy to Taylor dispersion in the ballistic regime \citep{taylor1922diffusion}. Indeed, we find that using $d = 0.06\tau_f u_\eta$ results in good agreement with our data across different $Re_\lambda$ and for different $r_p$, as shown 
by the dash-dotted lines in Fig.~\ref{fig:figure3}(a,b). This effect is only relevant at $St_p \lessapprox 0.1$ and becomes negligible once $r_c\gg d$ due to the inertial effect at larger $St_p$.

Another finding is that the incoming particle number density can differ from the global value. This is caused by clustering of bubbles and particles in different regions of the flow leading to segregation \citep{chan_bubbleparticle_2023}. As a result $\langle n_p'\rangle/n_p$ shown in Fig. ~\ref{fig:figure3}(d) decreases as $St_p$ increases reaching a minimum value of about 0.8 for $St_p \approx 1$, where clustering effects are known to be strongest. The segregation effect explains why the inertial limit $(1+r_p/r_b)^2$ at high $St_p$ is not reached in the simulations and in the model, 
where this effect is accounted for by multiplying with the factor $\langle n_p'\rangle/n_p$ obtained from the simulations.

\section{Conclusions}
We have elucidated the relevant mechanisms governing the collision rate of inertial particles with a finite-size bubble in turbulence. We demonstrated that for the investigated practical conditions, inertial effects induced by the flow around the bubble are the dominant effect. The non-linear dependence of these effects on the bubble slip velocity leads to an increase of the collision rate in turbulence of up to 100\% at $St_p \approx 0.1$ compared to quiescent flow. An additional increase in the turbulent collision rate is due to the short temporal correlation of the bubble slip velocity in the free-rising case. The effect of the resulting fluctuations during the bubble-particle interaction can be captured by an increase in the effective collision radius of the particle and is mostly relevant in the tracer limit for $St_p\lessapprox 0.1$. Segregation of bubbles and particles in turbulence reduces the particle density encountered by the bubble and and hence the collision rate by up to 20\% at $St_p\approx 1$. Remarkably, the effect of turbulence induced motion of the particles was found to have negligible impact during the transient bubble-particle interaction, whereas the incoming particle number density is reduced for particles with large Stokes number.
The developed frozen turbulence model provides a physically consistent framework that can be easily extended to a full collision model (by combining it with a prediction for $f(U_b')$). The approach is also transferable to other conditions, such as more complex shapes (by adopting a different parameterisation of $E_c$) and thus offers a more general relevance for collisions with finite-size objects in turbulence.

\begin{acknowledgments}
We thank Timothy Chan and Duco van Buuren for fruitful discussions. This project has received funding from the European Research Council (ERC) under the European Union’s Horizon 2020 research and innovation programme (grant agreement No. 950111, BU-PACT). This work was carried out on the Dutch national e-infrastructure with the support of SURF Cooperative.
\end{acknowledgments}
\textbf{Declaration of Interests.} The authors report no conflict of interest.

\appendix

\section{}\label{appA}
\begin{figure}
	\begin{center}
		\includegraphics[width=1\columnwidth]{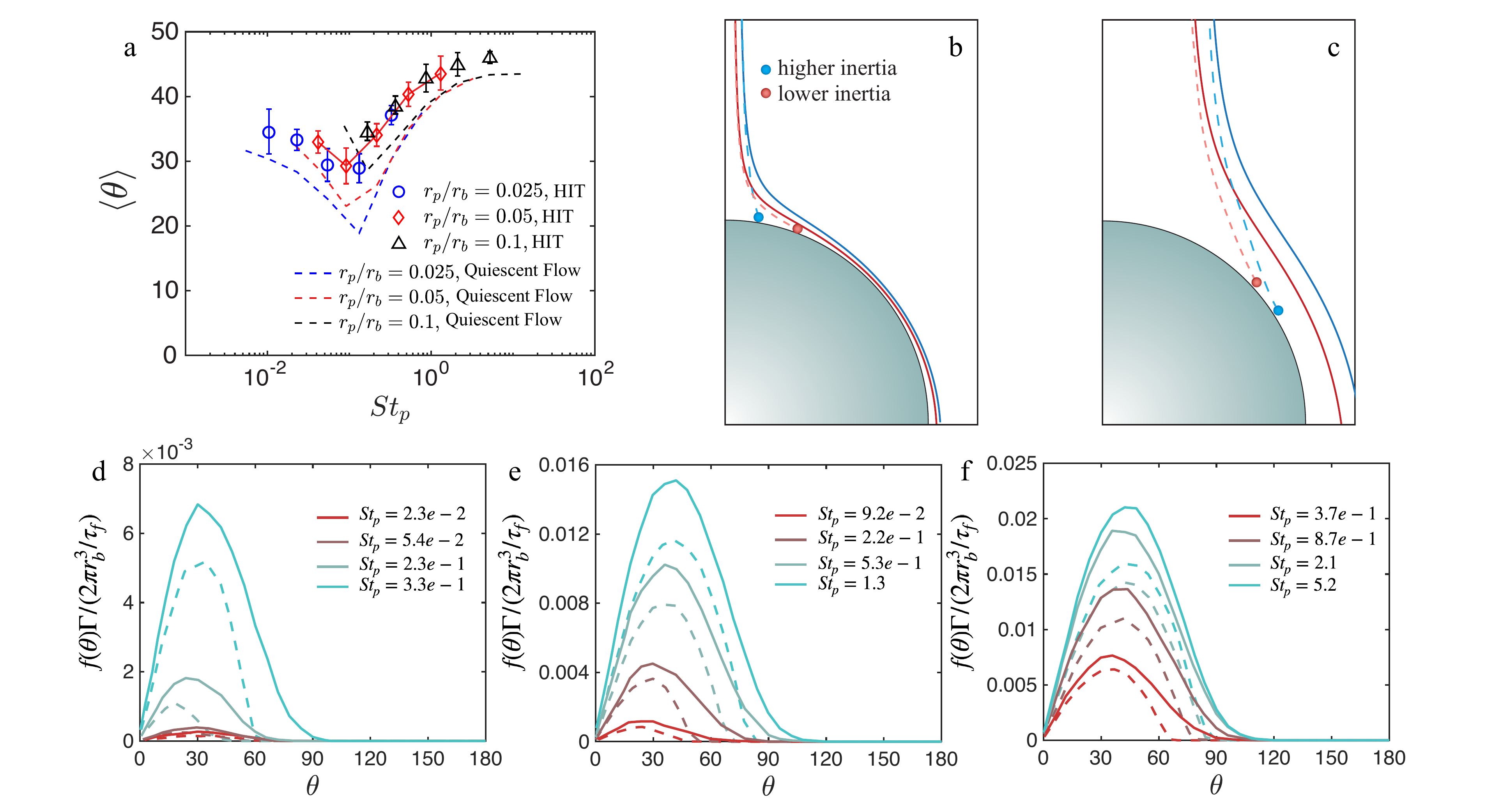}	
		\caption{(a) Mean collision angle $\langle\theta\rangle$ as a function of $St_p$. 
        Sketches in panel (b) and panel (c) illustrate the mechanism that $\langle\theta\rangle$ declines/increase as increasing inertia when $St_p$ is low/large in quiescent flow. The solid lines denote the streamlines which the particles originally stay and the dashed lines are the particle trajectories. Panel (d-f) show the PDF of collision angle $f(\theta)$ scaled by the normalized collision kernel for particle size $r_p/r_b=0.025,~0.05,~0.1$ in HIT (solid lines) and quiescent flow (dashed lines).
		}
		\label{fig:collision_angle}			
    \end{center}
\end{figure}

As we observe that the trajectories of collided particles in HIT are significantly different from those in quiescent flow, it is interesting to study how turbulence affects the collision angle. 
In Fig.~\ref{fig:collision_angle} we investigate the collision angle $\theta$ for the case where the bubble velocity is imposed. Here, $\theta$ denotes the angle between the direction of the imposed bubble velocity and the vector of collision position in the bubble frame. We observe that the mean collision angle $\langle \theta \rangle$ for the turbulence flow is higher than that for the case of quiescent flow. However, $\langle\theta\rangle$ shows a similar dependence on $St_p$ for both cases, where $\langle\theta\rangle$ first decreases as $St_p$ and rises after. This can be explained by inertial effects on the particles as illustrated in Fig.~\ref{fig:collision_angle} (b) and (c) for the deterministic case. At low $St_p$, $r_c$ is small. Consequently, particles collide on the bubble front region, where the streamlines are strongly curved. Therefore particles trajectories significantly deviate from the streamlines as $St_p$ increases. 
As a consequence, a higher inertia particles collide at a smaller collision angle, as illustrated in Fig.~\ref{fig:collision_angle} (b). 
When $St_p$ surpasses the critical value, $r_c$ becomes larger. In this case, increasing particle inertia leads to a higher collision rate as well as to a higher  collision angle, which is illustrated in Fig.~\ref{fig:collision_angle} (c).
These general trends for $\langle\theta\rangle$ are retained in the turbulent case, which is in line with our approach of representing the instantaneous collision process by the deterministic case.
Fig.~\ref{fig:collision_angle} (d-f) show the scaled PDF of collision angle $\theta$ for different particle size, respectively. The scaled PDF illustrates the distribution of the collision angle, as well as where the collision difference occur between HIT and quiescent flow. The distributions of collision angle indicate that the collisions are more likely to take place at intermediate angle. Moreover, we observe that the distribution extend to a wider range in HIT and more collisions occurs, which are consistent with the observations for $\langle\theta\rangle$.

\end{document}